\documentclass[aps,prb,onecolumn,superscriptaddress,10pt,twocolumn]{revtex4-2}

\usepackage{graphicx,url,hyperref,dcolumn,bm,multirow,amsmath,amssymb,amsfonts,subfigure,soul,wrapfig}
\usepackage[dvipsnames,svgnames,x11names]{xcolor}
\usepackage[symbol]{footmisc}
\usepackage[normalem]{ulem}
\usepackage[version=4]{mhchem}

\hypersetup{colorlinks,linkcolor=Red,urlcolor=OrangeRed,citecolor=Blue}

\usepackage[scaled]{helvet}

\usepackage{makecell}

\begin{document}
	
	\title{Single-ion anisotropy-stabilized short-period helimagnetism \\ in frustrated chiral \ce{Co5TeO8}}
	
	\author{Priya R. Baral}
	\email{priya.baral@epfl.ch}
	\affiliation{Department of Applied Physics and Quantum-Phase Electronics Center, The University of Tokyo, Bunkyo-ku, Tokyo 113-8656, Japan}
	\affiliation{PSI Center for Neutron and Muon Sciences, 5232 Villigen PSI, Switzerland}
	\affiliation{Institute of Physics, \'Ecole Polytechnique F\'ed\'erale de Lausanne (EPFL), CH-1015 Lausanne, Switzerland}
	
	\author{Ravi Yadav}
	\affiliation{Institute of Physics, \'Ecole Polytechnique F\'ed\'erale de Lausanne (EPFL), CH-1015 Lausanne, Switzerland}
	
	\author{Victor Ukleev}
	\affiliation{PSI Center for Neutron and Muon Sciences, 5232 Villigen PSI, Switzerland}
	\affiliation{Helmholtz-Zentrum Berlin f\"ur Materialien und Energie, D-14109 Berlin, Germany}
	
	\author{Thomas LaGrange}
	\affiliation{Institute of Physics, \'Ecole Polytechnique F\'ed\'erale de Lausanne (EPFL), CH-1015 Lausanne, Switzerland}
	
	\author{Ivica \v{Z}ivkovi\'{c}}
	\affiliation{Institute of Physics, \'Ecole Polytechnique F\'ed\'erale de Lausanne (EPFL), CH-1015 Lausanne, Switzerland}
	
	\author{Wen Hua Bi}
	\affiliation{Institute of Physics, \'Ecole Polytechnique F\'ed\'erale de Lausanne (EPFL), CH-1015 Lausanne, Switzerland}
	
	\author{Marek Bartkowiak}
	\affiliation{PSI Center for Neutron and Muon Sciences, 5232 Villigen PSI, Switzerland}
	
	\author{Robert Cubitt}
	\affiliation{Institut Laue–Langevin, 71 avenue des Martyrs, CS 20156, Grenoble, 38042 Cedex 9, France}
	
	\author{Nina-Juliane Steinke}
	\affiliation{Institut Laue–Langevin, 71 avenue des Martyrs, CS 20156, Grenoble, 38042 Cedex 9, France}
	
	\author{Vladimir Pomjakushin}
	\affiliation{PSI Center for Neutron and Muon Sciences, 5232 Villigen PSI, Switzerland}
	
	\author{Yurii Skourski}
	\affiliation{Dresden High Magnetic Field Laboratory (HLD-EMFL), Helmholtz-Zentrum Dresden-Rossendorf, 01328 Dresden, Germany}
	
	\author{Henrik M. R\o nnow}
	\affiliation{Institute of Physics, \'Ecole Polytechnique F\'ed\'erale de Lausanne (EPFL), CH-1015 Lausanne, Switzerland}
	
	\author{Oleg V. Yazyev}
	\affiliation{Institute of Physics, \'Ecole Polytechnique F\'ed\'erale de Lausanne (EPFL), CH-1015 Lausanne, Switzerland}
	
	\author{Arnaud Magrez}
	\affiliation{Institute of Physics, \'Ecole Polytechnique F\'ed\'erale de Lausanne (EPFL), CH-1015 Lausanne, Switzerland}
	
	\author{Jonathan S. White}
	\email{jonathan.white@psi.ch}
	\affiliation{PSI Center for Neutron and Muon Sciences, 5232 Villigen PSI, Switzerland}
	
	\date{\today}
	
	\maketitle
	
	\newpage

	\begin{center}
		\textbf{ABSTRACT}
	\end{center}
	
	\noindent \textbf{Chiral spin textures in magnetic insulators promise magneto-electric (ME) spintronics with orders-of-magnitude lower power consumption than metallic systems. However, realizing the short magnetic periods required for high-density device integration remains difficult, as conventional Dzyaloshinskii-Moriya interaction (DMI)-based mechanisms typically constrain spiral periods to tens of nanometers. While theory predicts that strong single-ion anisotropy (SIA) on frustrated lattices can stabilize complex non-coplanar textures, the potential for using this mechanism to engineer such compact textures remains largely unexplored. Here we report that a cubic chiral insulator \ce{Co5TeO8} provides an experimental example of this paradigm. Comprehensive neutron scattering and magnetometry reveal helimagnetic spirals with continuously tunable pitch of 5.7–10~nm embedded in a complex phase diagram spanning eight distinct phases. Capacitance anomalies throughout the phase diagram indicate magneto-electric coupling, pointing to the possibility of future $E$-field control of these textures. The temperature- and field-dependence of the helical wavevector strongly support a scenario in which site-dependent SIA provides the leading contribution to the selection of the helical period from a frustration-induced degenerate manifold. Consistent with this interpretation, \textit{ab initio} calculations place SIA approximately an order of magnitude above DMI, distinct from conventional helimagnets. \ce{Co5TeO8} thus offers an experimental realization of sub-10~nm helimagnetism and motivates a design principle for anisotropy-engineered correlated insulators.}
	
	
	\begin{center}
		\textbf{MAIN}
	\end{center}
	
	\noindent\textbf{Introduction.}
	
	\noindent Non-coplanar magnetic states---ranging from topological skyrmions to chiral vortex lattices---are central to modern spintronics owing to their emergent electrodynamics and robust stability against weak perturbations~\cite{fert2017magnetic,matsui2021skyrmion,hamamoto2015quantized,kurumaji2019skyrmion}. Yet, these intricate multi-$\bm{Q}$ architectures typically nucleate from a more fundamental parent phase: the single-$\bm{Q}$ spiral state~\cite{ghimire2020competing,baral2025fluctuation}. As the elementary building block of non-coplanar magnetic states, the single-$\bm{Q}$ spiral emerges when competing interactions preclude collinear spin alignment~\cite{tokura2010multiferroics,tokura2014multiferroics}, making a microscopic understanding of its stability essential for engineering complex topological phases~\cite{nagaosa2013topological,tokura2020magnetic}. Conventionally, this stability is attributed to specific exchange pathways governed by crystal symmetry and electronic structure of the material: the antisymmetric Dzyaloshinskii-Moriya interaction (DMI) in chiral non-centrosymmetric hosts~\cite{dzyaloshinsky1958thermodynamic,moriya1960anisotropic,bak1980theory}, geometric frustration~\cite{richter2010frustrated,okubo2012multiple} or Fermi-surface-nested Ruderman-Kittel-Kasuya-Yosida (RKKY) interactions in centrosymmetric systems~\cite{bouaziz2022fermi,nomoto2020formation}, and, in itinerant magnets, higher-order terms such as the biquadratic exchange interaction~\cite{hayami2017effective,batista2016frustration}. Crucially, while these diverse mechanisms successfully predict a wide array of non-collinear magnetic phases, they share a common paradigm in which spatial modulation is driven primarily by inter-site exchange, with local energy scales typically treated as subleading.
	
	\begin{figure*}[htb!]
		\centering
		\includegraphics[width=0.95\linewidth]{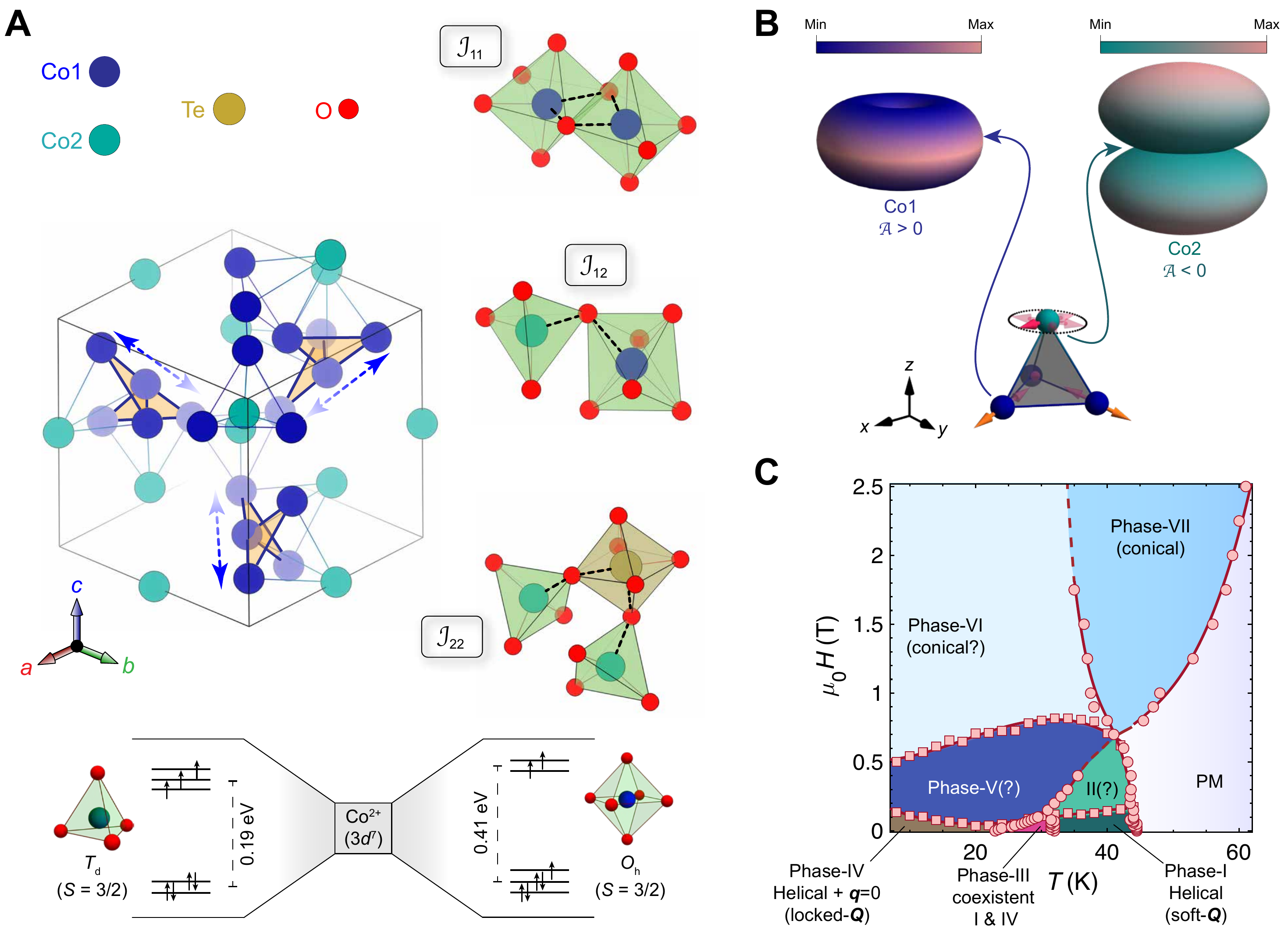}
		\caption{\textbf{Phase diagram and local magnetic anisotropy of chiral cubic \ce{Co5TeO8}.} (A) Crystal structure of \ce{Co5TeO8} viewed slightly away from [111] axis, comprising of two inequivalent Co ions: Co1 in blue and Co2 in teal. For simplicity, Te and O atoms are omitted from the full unit cell depiction. The dashed double arrows represent the Co1 ions arranged along $\left< 110\right>$, which form triangular plaquettes (shaded yellow). Two corner-sharing plaquettes are mutually rotated by 110$^{\circ}$ with respect to each other. Ultimately, they form corner-sharing tetrahedra with Co2 ions. The connectivity of various Co-O polyhedra are shown, with nearest-neighbour (NN) and next-nearest-neighbour (NNN) superexchange pathways represented by dashed lines. Here, $\mathcal{J}_{11}$, $\mathcal{J}_{12}$, and $\mathcal{J}_{22}$ represents bilinear exchanges between Co1-Co1, Co1-Co2, and Co2-Co2 sites, respectively. Crystal electric field splitting diagrams of Co$^{2+}$ ions in tetrahedral and octahedral environments are depicted. (B) \textit{Ab initio} calculations indicate contrasting single-ion anisotropies ($\mathcal{A}$), identifying the Co1 sublattice as having easy-axis character and the Co2 sublattice as easy-plane. Consequently, the schematic depicts Co1 spins pointed along the local $\langle\text{111}\rangle$ axes, with Co2 spins constrained within the plane orthogonal to the local $z$-axis. (C) Magnetic phase diagram of \ce{Co5TeO8} constructed from high-resolution $M(H)$ and $\chi_{\mathrm{ac}}(T)$ data. We identify up to 8 distinct phases in the explored parameter space. Filled circles and rectangles represent phase boundaries extracted from $\chi_{\mathrm{ac}}(T)$ and \textit{M}(\textit{H}) data, respectively. These data can be found in Fig.~\ref{Fig2} as well as Supplementary Materials Fig.~{\color{Red}S5} and Fig.~{\color{Red}S6}. For magnetic field dependent phase boundaries, we only consider data obtained while the field is ramped up. Small-angle neutron scattering analysis indicates that the helical wavevector is locked at $\bm{Q} = 1.1(1)$~nm$^{-1}$ in Phase-IV, whereas it exhibits a monotonic change of its magnitude in Phase-I (``soft-$\bm{Q}$''). In Phase-II, the scattering intensity is distributed along an arc with constant angular width and linearly decaying intensity, eventually evolving into the characteristic two-spot pattern of the conical state (Phase-VII). See Fig.~\ref{Fig5} and main text for further details.
		}
		\label{Fig1}
	\end{figure*}
	
	Conversely, magnetic anisotropy—traditionally treated as a weak perturbative correction—can fundamentally restructure the magnetic energy landscape when its magnitude rivals the exchange energy scales~\cite{hayami2016bubble,hayami2021noncoplanar,lin2015skyrmion}. In isotropic Heisenberg models, single-$\bm{Q}$ helices are generically favored as a simple harmonic modulation preserves the fixed-moment constraint, whereas multi-$\bm{Q}$ superpositions tend to generate amplitude modulations that require higher harmonics to restore normalization. However, this exchange cost can be offset by strong easy-axis single-ion anisotropy (SIA), which, aided by thermal softening of the longitudinal spin stiffness~\cite{hayami2016bubble}, promotes multi-$\bm{Q}$ formation. On frustrated lattices, this mechanism stabilizes a hierarchy of exotic phases, including the skyrmion lattice~\cite{leonov2015multiply}. Moreover, in multi-sublattice magnets, inequivalent SIA tensors can compete with the leading exchange to promote noncollinearity~\cite{rinaldi1979model}, yet experimental realizations of anisotropy-dominated helimagnets remain elusive.
	
	Here, we combine comprehensive experimental characterization with first-principles calculations to investigate helimagnetism in chiral \ce{Co5TeO8}. Our density functional theory calculation predicted a 127 meV charge-transfer gap, motivating the synthesis and investigation of its magnetic properties. Using magnetic and thermodynamic probes, we constructed a high-resolution multi-phase diagram consisting of eight different phases. In addition, we employ neutron scattering, high-field magnetometry, and capacitance measurements to map this magnetic landscape and characterize compact field-tunable helical textures. These experiments indicate that site-dependent SIA, rather than DMI, is the leading interaction governing the stabilization of the helical state---a scenario further supported by \textit{ab initio} many-body wavefunction calculations. Together with the experimental observations, these results suggest a framework for engineering compact magnetic textures in correlated oxides, with potential implications for magneto-electric (ME) spintronics.
	
	
	\vspace{10pt}
	
	\noindent\textbf{Strong electronic correlations in \ce{Co5TeO8}.}
	
	\noindent To search for magnets with enhanced SIA, we turned our attention to Co-based systems, targeting spinel-type \ce{Co5TeO8}. Although its magnetic ground state depends on structural polymorphism~\cite{podchezertsev2021influence}, we focused on the chiral cubic phase ($P4_132$/$P4_332$), where geometric frustration and broken inversion symmetry enable complex magnetic textures. The combination of these structural features with \ce{Co5TeO8}’s partially filled $d$-orbitals, likely involving strong electronic correlations, motivated us to investigate its electronic structure.
	
	To assess the metallic versus insulating character of its electronic structure, we performed DFT calculations within the LDA+$U$ framework, using the experimental crystal structure (see Fig.~\ref{Fig1}A). The LDA+$U$ approach, incorporating on-site Coulomb repulsion via a Hubbard $U$ parameter, has successfully reproduced band gaps in Mott insulators such as \ce{Na2IrO3}~\cite{comin2012novel}. For \ce{Co5TeO8}, while standard LDA predicts metallic behavior, introducing $U$ = 3~eV opens a charge-transfer gap of 127~meV at the $\Gamma$-point (Fig.~{\color{Red}S1}), a result robust to the choice of $U$. The predominant $d$-character of the bands near the Fermi energy indicates strong electronic correlations, implying sensitivity to external fields and potential ME coupling.
	
	To explore magnetism of \ce{Co5TeO8}, we synthesized a high-quality microcrystalline sample and performed comprehensive structural characterization. Structural analysis confirms that \ce{Co5TeO8} crystallizes in the chiral space group $P4_332$ (or its enantiomorph $P4_132$), with both inequivalent magnetic Co-sites adopting 2+ oxidation state as confirmed from electron energy loss spectroscopy measurements (Supplementary Materials Section~{\color{Red}S2}). Further structural details are provided in Supplementary Materials Section~{\color{Red}S1}. The Co1 sublattice (12$d$ sites) forms corner-sharing triangular motifs that interconnect with Co2 ions (8$c$ sites) to complete a three-dimensional framework of corner-sharing tetrahedra. This geometry generates strong magnetic frustration where competing exchange interactions between the two sublattices may suppress conventional magnetic order~\cite{ukleev2021frustration}. The combination of strong electronic correlations on the chiral frustrated lattice establishes \ce{Co5TeO8} as a platform for hosting nontrivial magnetic textures. As we demonstrate in the following sections, localized 3$d$ electrons in \ce{Co5TeO8} enable tunable magnetic phenomena with a rich phase diagram.
	
	\vspace{10pt}
	
	\noindent\textbf{Complex field-tunable magnetic phase diagram.}
	
	\begin{figure}[htb!]
		\centering
		\includegraphics[width=1.0\linewidth]{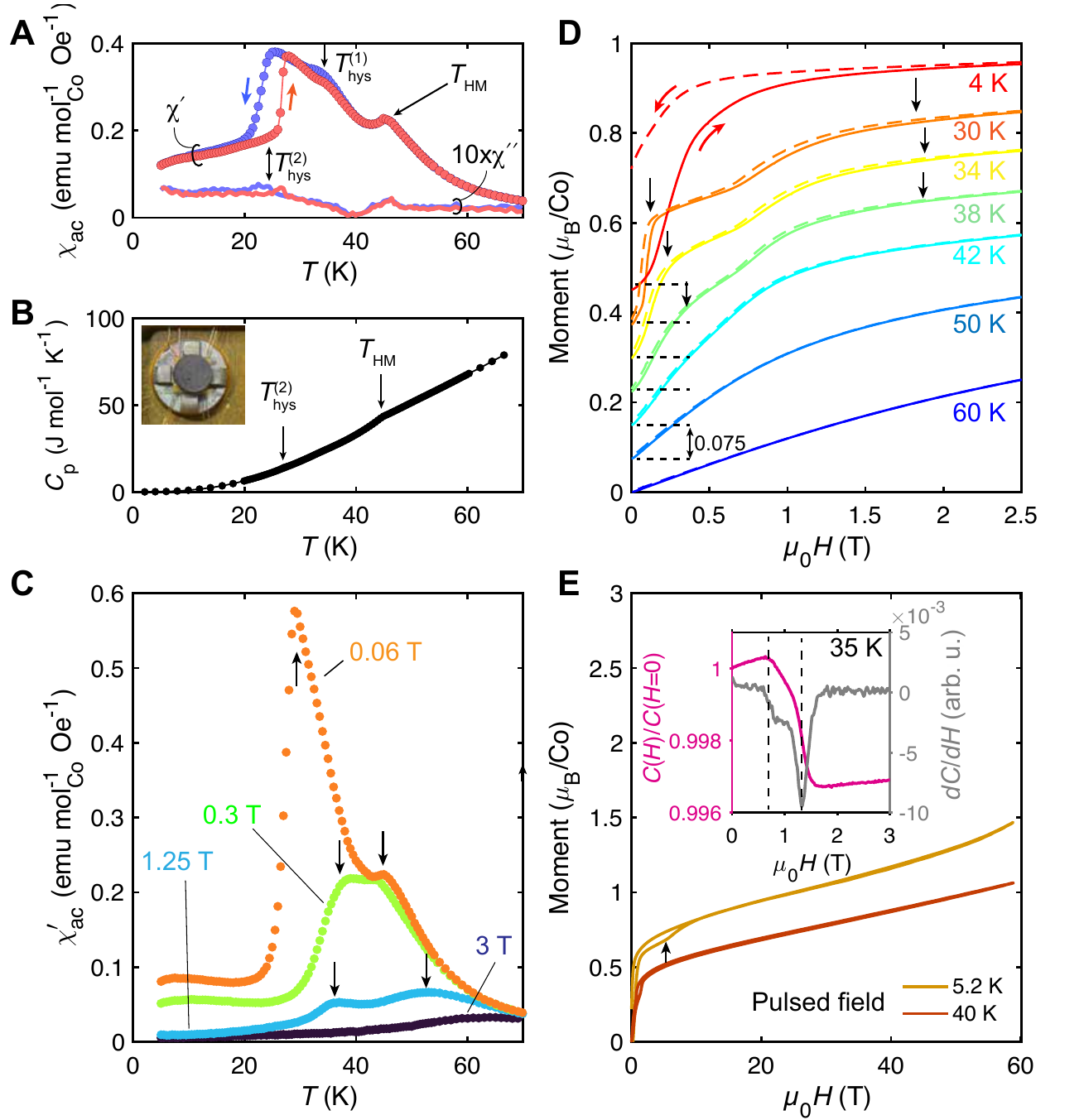}
		\caption{\textbf{Magnetic and thermodynamic response across the multi-phase diagram of magneto-electric \ce{Co5TeO8}.} (A) Shows the result of ac susceptibility measurement on \ce{Co5TeO8} polycrystalline powder in zero field. A constant factor of 10 has been applied to the imaginary part of the total susceptibility ($\chi^{\prime\prime}_{\mathrm{ac}}$) for better visualization. (B) Specific heat data collected in zero field shows a weak anomaly at $T_{\mathrm{HM}}$. Inset shows the pressed pellet ($\phi\sim$3.2~mm) used for these measurements. (C) Ac susceptibility data as a function of temperature measured at four different magnetic fields, scanning across various transitions. (D) Shows magnetization evolution as a function of magnetic field at seven different temperatures selected according to the phase diagram shown in Fig.~\ref{Fig1}C. (E) High pulsed magnetic field data of \ce{Co5TeO8} collected at two temperatures show no sign of saturation magnetization up to 60~T. Inset shows evolution of capacitance ($C$), and the corresponding $dC/dH$, as a function of external magnetic field at a fixed temperature of 35~K. The dashed vertical lines are the phase boundaries extracted earlier. Black arrows in all panels mark the occurrence of temperature (or magnetic field)-induced transitions.
		}
		\label{Fig2}
	\end{figure}
	
	\noindent High-resolution magnetometry reveals a complex magnetic energy landscape in \ce{Co5TeO8} with multiple phase transitions, as shown in Fig.~\ref{Fig1}C. The real component of ac susceptibility ($\chi^{\prime}_{\mathrm{ac}}$) exhibits a sharp peak at 44.9~K ($T_{\mathrm{HM}}$), indicating onset of long-range magnetic order in Phase-I (see Fig.~\ref{Fig2}A). This transition displays clear dissipative character, as evidenced by a corresponding peak in imaginary susceptibility ($\chi^{\prime\prime}_{\mathrm{ac}}$). Below $T_{\mathrm{HM}}$, two additional hysteretic regions emerge at $T^{(1)}_{\mathrm{hys}}$ and $T^{(2)}_{\mathrm{hys}}$, consistently observed in both ac susceptibility and dc magnetization. Applied magnetic fields cause these transitions to converge and shift toward $T_{\mathrm{HM}}$ (Fig.~\ref{Fig2}C and Supplementary Materials Section~{\color{Red}S3}), suggesting significant renormalization effects similar to skyrmion lattice hosts~\cite{zivkovic2014scaling,janoschek2013fluctuation,kindervater2019weak}.
	
	Heat capacity measurements reveal contrasting thermodynamic signatures. While magnetometry shows pronounced features at all three transitions, heat capacity exhibits only weak anomalies at $T_{\mathrm{HM}}$ and $T^{(2)}_{\mathrm{hys}}$, with no discernible signature at $T^{(1)}_{\mathrm{hys}}$ (see Fig.~\ref{Fig2}B). This suppressed response despite clear magnetic signatures suggests minimal entropy changes governed by persistent short-range correlations, previously identified in Ref.~\cite{podchezertsev2021influence}. Field-dependent magnetometry data reveal pronounced metamagnetic transitions below $T_{\mathrm{HM}}$, becoming increasingly steeper at lower temperatures with broadened hysteresis, as shown in Fig.~\ref{Fig2}D. These field-driven responses coincide with sharp anomalies in electrical capacitance (Fig.~\ref{Fig2}E inset and Supplementary Materials Section~{\color{Red}S3}), confirming ME coupling in \ce{Co5TeO8} crucial for functional control. Remarkably, magnetic saturation is absent even at 60~T pulsed fields (Fig.~\ref{Fig2}E), presumably reflecting frustration in the strongly-coupled tetrahedral framework where competing exchanges prevent forced ferromagnetic alignment~\cite{janson2014quantum}. This intricate behavior necessitates comprehensive neutron scattering to resolve the underlying magnetic structure.
	
	\begin{figure*}[htb!]
		\centering
		\includegraphics[width=0.95\linewidth]{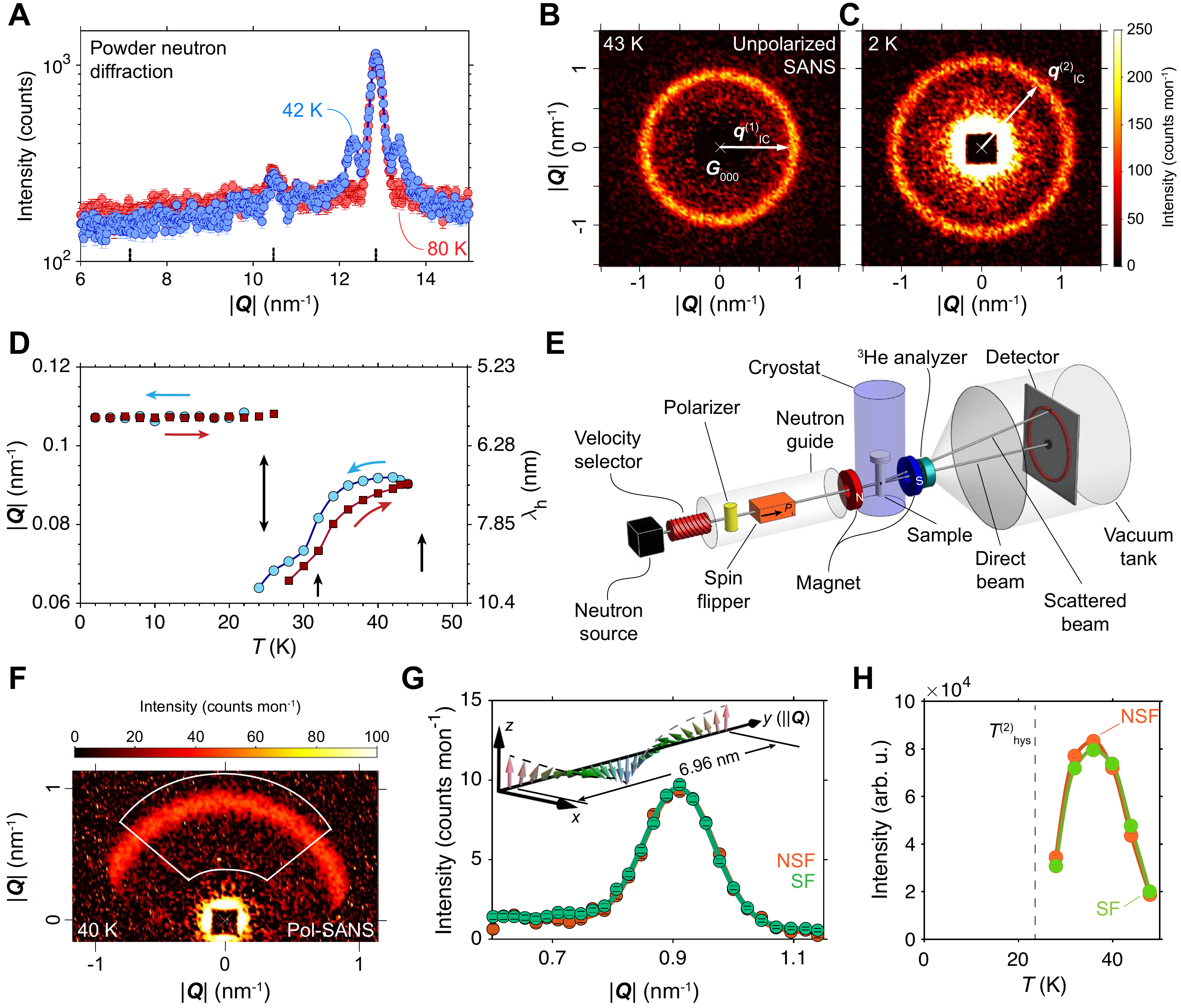}
		\caption{\textbf{Compact helimagnetic order revealed by neutron scattering.} (A) Wide angle neutron diffractograms of \ce{Co5TeO8} obtained at two temperatures above and below $T_{\mathrm{HM}}$, focused on two reflections of the principal cubic directions show appearance of magnetic satellites around them. The nuclear reflections (from low- to high-$\bm{Q}$: (100), (110) and (111)) are marked by dashed vertical lines. We do not observe any incommensurate (IC) satellites around forbidden (001) reflection. (B, C) Show the SANS 2D detector image obtained just below $T_{\mathrm{HM}}$ and at \textit{T} = 2~K, respectively. The circular distribution of the magnetic scattering intensity is consistent with the expectation for magnetic scattering arising at propagation vector $\bm{q}$ within randomly oriented grains of the polycrystalline sample. (D) Shows the length of incommensurate magnetic wavevectors (and the corresponding pitch length of the spiral in real space, $\lambda_{\mathrm{h}}$) as a function of temperature. Phase boundaries, as obtained from our $\chi_{\mathrm{ac}}$ data, have been clearly marked by vertical arrows. (E) Schematic of the uniaxial polarization-analysis setup employed at D33 to investigate the IC state in \ce{Co5TeO8}. A uniform longitudinal guide field was applied along the incoming beam direction in order to maintain the polarization of the incoming beam. Representative detector image is also sketched, with a characteristic circular magnetic scattering pattern. The setup for unpolarized SANS measurements is similar, but with polarizer, spin flipper and analyzer removed. (F) Shows a representative detector image obtained at $T~=~40~$K. Polarization analysis is only performed on the scattered neutrons forming the upper arc on the detector (inside the white sector box). Due to the analyzer's finite size, only scattering within the top portion of the detector could be analyzed. (G) Integrated intensity of non spin-flip and spin-flip polarized SANS signals observed in Phase-I at 40~K. An equal distribution of NSF scattering intensity combined with SF scattering is consistent with a Bloch-type modulation. Inset shows a real-space sketch of the proper helical magnetic structure realized in \ce{Co5TeO8}. The indicated pitch length of the helix is the one obtained at 44~K, just below $T_{\mathrm{HM}}$. (H) Shows evolution of NSF-SF scattering intensity between $T_\text{HM}$ and $T^{(1)}_\text{hys}$.
		}
		\label{Fig3}
	\end{figure*}
	
	\vspace{10pt}
	
	\noindent\textbf{Compact spiral order revealed by neutron scattering.} 
	
	\noindent To uncover the magnetic structure in \ce{Co5TeO8}, we performed powder neutron diffraction (PND) measurements. Just below $T_{\text{HM}}$, magnetic satellites emerge around ($h$,$h$,$h$) and ($h$,$h$,0) nuclear reflections (see Fig.~\ref{Fig3}A), characterized by incommensurate propagation vectors $\bm{Q}_{\text{IC}}^{\text{1a}} = \bm{G}_{111} \pm \bm{q}^{(1)} = \bm{G}_{110} \pm \bm{q}^{(1)}$, consistent with a previous report~\cite{podchezertsev2021influence}. Notably, no magnetic satellites appear around $\bm{G}_{h00}$ reflections within instrumental resolution. The systematic absences of ($h$,0,0) reflections ($h \neq 4n$) confirm that the $4_1$ (or $4_3$) screw symmetry of the parent space group is retained, though overall cubic magnetic symmetry is broken by incommensurate modulation.
	
	Small-angle neutron scattering (SANS) measurements provide complementary information about the magnetic order. Due to powder averaging, the discrete incommensurate $\bm{Q}$-vectors manifest as ring-like scattering patterns on the two-dimensional multidetector. Figure~\ref{Fig3}B and \ref{Fig3}C show data at 44~K and 2~K, respectively. The ring-like scattering centered at the $\Gamma$-point corresponds to $\bm{Q}_{\text{IC}}^{\text{(1)}}$ with $|\bm{Q}|$ = $0.903(2)$~nm$^{-1}$ at $T_\text{HM}$ and $\bm{Q}_{\text{IC}}^{\text{(2)}}$ with $|\bm{Q}|$ = $1.1\pm0.1$~nm$^{-1}$ below $T^{(2)}_{\text{hys}}$ (Fig.~\ref{Fig3}D). The corresponding pitch length at 2~K is $\lambda_{\text{h}} = 5.71\pm0.51$~nm, increasing with temperature from $\lambda_\mathrm{h}$($T_{\text{HM}}$) = $6.96\pm0.02$~nm to $\lambda_\mathrm{h}$($T^{(1)}_{\mathrm{hys}}$) = $9.66\pm0.15$~nm. These compact magnetic periodicities are among the shortest reported in chiral insulator magnets. See Table~\ref{tab:TopoList} for a thorough comparative list of most-relevant chiral magnets. However, the unpolarized neutron scattering measurements alone cannot distinguish between different spiral helicities. To determine whether these ultracompact spirals adopt proper-screw or cycloidal character, we performed polarization-analyzed SANS measurements as described in the following section.
	
	\vspace{10pt}
	
	\noindent\textbf{Bloch type modulation deduced from polarized SANS.} 
	
	\noindent Having established the compact spiral order, we performed uniaxial polarization-analyzed SANS to determine the helical spin texture. A schematic of the experimental setup is shown in Fig.~\ref{Fig3}E. The incoming neutron polarization vector is aligned (or anti-aligned) with the neutron beam. Non-spin-flip (NSF) and spin-flip (SF) scattering cross-sections are probed using a spin state analyzer placed between sample and detector. This configuration enables the decomposition of the magnetic interaction vector $m_{\perp}(\bm{Q})$ into components defined by the polarization axis. The NSF cross-section isolates the projection of $m_{\perp}(\bm{Q})$ parallel to the initial polarization $P_i$ (as indicated in Fig.~3E), whereas the SF channel is sensitive to the component simultaneously orthogonal to both $P_i$ and $\bm{Q}$. Consequently, for a helimagnetic spiral characterized by moments distributed isotropically in the plane normal to $\bm{Q}$, the scattering intensity is expected to be equally distributed between the NSF and SF channels. In contrast, for an easy-plane cycloidal structure with moments confined to the $\bm{Q}$-$x$ (or $\bm{Q}$-$z$) plane, the magnetic scattering is expected to appear exclusively in the NSF (or SF) channel, respectively. See Ref.~\cite{kurumaji2021direct,takagi2022square} for more details. Since the ring-like scattering pattern (Fig.~\ref{Fig3}F) represents a superposition of multiple magnetic domains, each angular segment provides equivalent information about NSF and SF cross-sections, enabling systematic analysis across different $\bm{Q}$-vector orientations.
	
	Analysis of polarized SANS data at 40~K shows nearly equivalent intensities in both NSF and SF channels for Phase-I (Fig.~\ref{Fig3}G). This balanced contribution indicates that magnetic Fourier components lie predominantly in the plane perpendicular to $\bm{Q}$ with comparable amplitudes in both horizontal and vertical directions, consistent with helimagnetic spiral arrangement (schematic shown in inset of Fig.~\ref{Fig3}G). The helicity is retained down to $T_{\text{hys}}^{(2)}$ as deduced from equal NSF-SF intensities (Fig.~\ref{Fig3}H). More detailed analysis using narrower sector boxes reproduces the same result (Supplementary Materials Section~{\color{Red}S4}). While helical modulation is expected due to the chiral crystal structure, the unusually short magnetic period sharply differentiates \ce{Co5TeO8} from archetypal DM-stabilized helimagnets~\cite{adams2011long,adams2012long,seki2012observation}. Having confirmed the Bloch-type helical character, we next investigate how applied magnetic fields modify the helical pitch and induce phase transitions, as revealed by field-dependent SANS measurements.
	
	\vspace{10pt}
	
	\noindent\textbf{Field-driven transformation of the helimagnetic spiral} 
	
\begin{figure*}[htb!]
		\centering
		\includegraphics[width=0.90\linewidth]{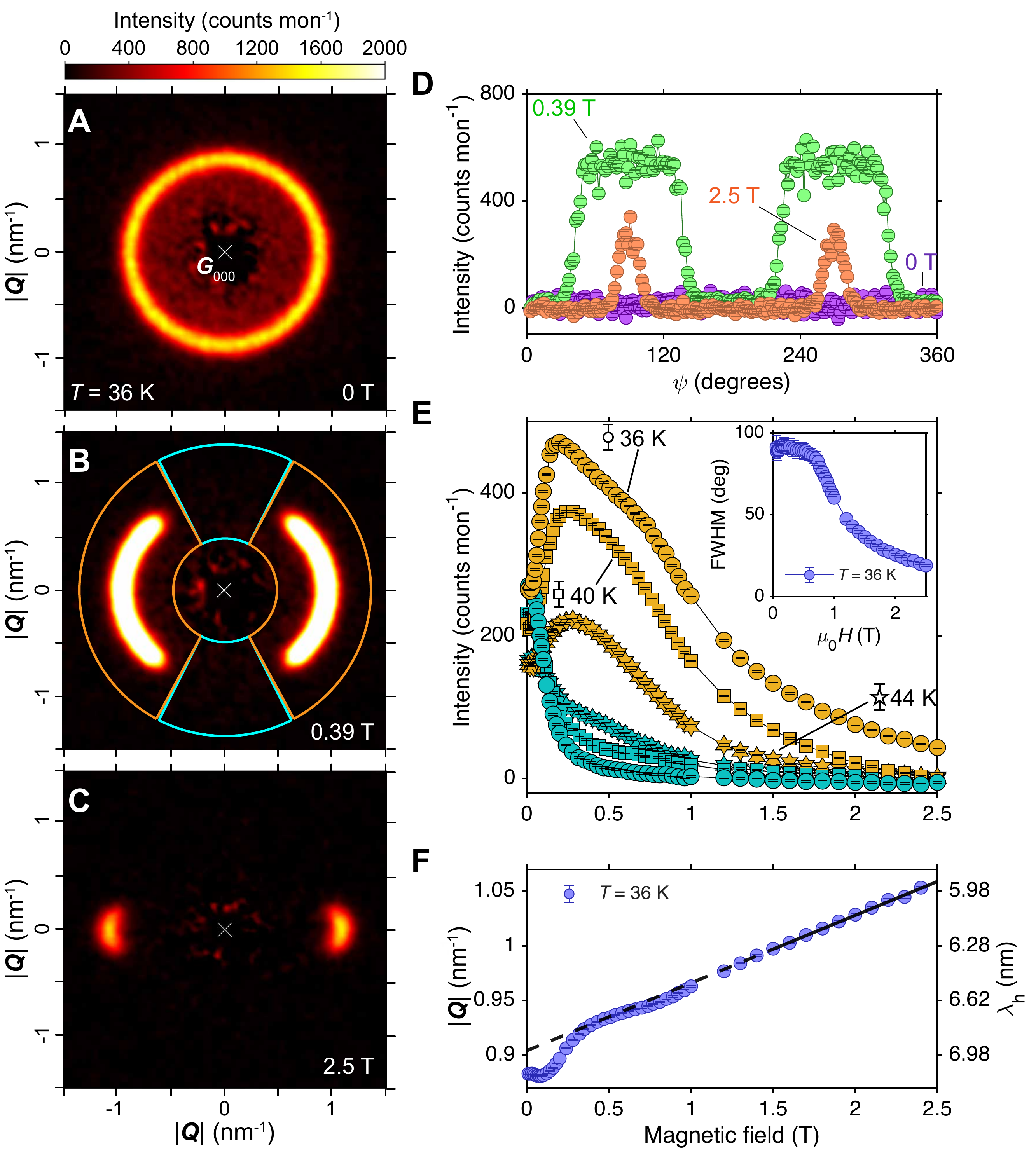}
		\caption{\textbf{Evolution of the helimagnetic spiral in transverse magnetic field.} (A)-(C) Detector images obtained in a magnetic field of 0~T, 0.39~T, and 2.5~T at $T~=~36~$K. The incident neutron beam is directed perpendicular to the detector plane, while the magnetic field is applied transverse to it (horizontal in the detector images). (D) Shows the corresponding azimuthal scans at the three magnetic field values. For clarity, a constant offset of -250 counts/mon is applied to the data at 0~T. Panel-(E) shows the magnetic field dependence of the scattering intensities extracted from the orange and cyan sector boxes indicated in Panel-B. The inset displays the full width at half maximum (FWHM) of the azimuthal intensity arcs as a function of magnetic field at $T$ = 36 K. (F) Field evolution of the incommensurate component of the magnetic order ($\bm{Q}$), together with its corresponding pitch length in real space. Data above 1.5~T are fitted to illustrate the linear scaling of $|\bm{Q}^{(1)}(H)|$ and the corresponding $\lambda_{\mathrm{h}}$.
		}
		\label{Fig5}
	\end{figure*}

	\noindent To investigate the magnetic field response of the incommensurate phases, we systematically tracked the evolution of $\bm{Q}^{(1)}_{\mathrm{IC}}$ under applied transverse magnetic fields ($\mu_{0}H \perp \bm{k}_i$) using SANS. Starting from zero-field ring-like scattering at 36~K in Phase-I (Fig.~\ref{Fig5}A), applied magnetic field gradually transforms the ring into an intermediate phase characterized by arcs of scattering (Fig.~\ref{Fig5}B) before entering a well-defined conical phase at higher fields (Fig.~\ref{Fig5}C). The conical phase is clearly identified by a pair of spots with propagation vectors aligned parallel and antiparallel to the applied field, consistent with other cubic helimagnets~\cite{adams2012long,adams2011long}. Further field increase suppresses incommensurate satellites above 2.5~T, leading to field-polarized magnetic order. Notably, the absence of scattering intensity in sectors perpendicular to the applied field (around $\psi$ = 0$^\circ$ or 180$^\circ$ in Fig.~\ref{Fig5}D and teal sector box in Fig.~\ref{Fig5}E) definitively excludes formation of the conventional multi-$\bm{Q}$ skyrmion lattice observed in other chiral cubic magnets.
	
	The intermediate phase (Phase-II) exhibits distinctive characteristics revealing complexity of underlying magnetic interactions. Unlike the conical phase with monotonic peak-width decay, the intermediate phase maintains nearly constant azimuthal intensity distribution across its stability range (Fig.~\ref{Fig5}E inset). This behavior suggests stabilization of a multi-domain incommensurate structure with $\bm{Q}$-vectors distributed between $\langle110\rangle$ and $\langle111\rangle$ directions, with the field only partially selecting preferred orientations. Remarkably, the incommensurate wavevector magnitude $|\bm{Q}^{(1)}|$ increases systematically with applied field (see Fig.~\ref{Fig5}F), indicating that the spiral shortens under field rather than collapsing into ferromagnetic-type alignment. This field-driven reduction of helical pitch is rare among DMI-stabilized helimagnets, necessitating \textit{ab initio} wavefunction-based calculations to understand microscopic magnetic interactions.
	
	\vspace{10pt}
	
	\noindent\textbf{Single-ion anisotropy-dominated magnetic energy landscape.} 
	
	\noindent The Co-O coordination geometries in \ce{Co5TeO8} deviate significantly from ideal octahedral and tetrahedral environments with $\angle$Co1-O-Co1 and $\angle$Co2-O-Co2 bond angles of 94.51(1)$^{\circ}$ and 116.77(1)$^{\circ}$, respectively. These distortions make prediction of exchange interactions via Goodenough-Kanamori-Anderson rules challenging, thus requiring explicit quantum chemistry calculations. Hence, we employed electrostatically embedded finite-size clusters extracted from the experimental crystal structure (see Fig.~\ref{Fig1}A), an approach well-suited for strongly correlated $d$-electron systems with highly localized electronic states~\cite{moreira1999local,bogdanov2011abinitio,katukuri2012ab,bogdanov2013magnetic,de2006unified,malrieu2014magnetic}. We used the following spin Hamiltonian encompassing both on-site and inter-site magnetic interactions:
	
	\begin{align}
		\mathcal{H} = \sum_{\langle i,j\rangle = 1,2}\mathcal{J}_{ij}(\mathcal{S}_i &\cdot \mathcal{S}_j) + \mathcal{K}_{ij}(\mathcal{S}_i \cdot \mathcal{S}_j)^2 + \mathcal{D}_{ij} \cdot \left(\mathcal{S}_i \times \mathcal{S}_j\right) \nonumber\\
		&+ \sum_{i}\mathcal{A}_i(\mathcal{S}_{i}\cdot \hat{e}_i)^2.
		\label{Eqn:Eq1}
	\end{align}
	
	\noindent The summation ${\langle i,j\rangle}$ runs over all nearest-neighbor bonds. The exchange parameters: bilinear ($\mathcal{J}_{ij}$), biquadratic ($\mathcal{K}_{ij}$), and antisymmetric DM interaction vectors ($\mathcal{D}_{ij}$~\cite{dzyaloshinsky1958thermodynamic,moriya1960anisotropic}) are bond-type-dependent, taking three distinct sets of values corresponding to Co1-Co1 ($\mathcal{J}_{11}, \mathcal{K}_{11}, \mathcal{D}_{11}$), Co2-Co2 ($\mathcal{J}_{22}, \mathcal{K}_{22}, \mathcal{D}_{22}$), and Co1-Co2 ($\mathcal{J}_{12}, \mathcal{K}_{12}, \mathcal{D}_{12}$) bonds, as shown in Fig.~\ref{Fig1}A. $\mathcal{A}_i$ is the SIA coefficient for site $i$, where $\hat{e}_i$ represents site-dependent anisotropy axis. Further details about our quantum chemistry calculations can be found in the Methods section, and Supplementary Materials Section~{\color{Red}S5}.
	
	We find that all three bilinear exchange pathways are antiferromagnetic with $\mathcal{J}_{11} = 0.116$~meV, $\mathcal{J}_{22} = 0.088$~meV, and $\mathcal{J}_{12} = 0.208$~meV (see Supplementary Materials Table~S2), at least 20 times weaker than the thermal energy scale ($k_\text{B}T_\text{HM} \sim 3.9$~meV) associated with $T_\text{HM}$, suggesting that the bilinear exchanges alone underestimate the long-range ordering temperature. The biquadratic interactions, in turn, are ferromagnetic-type with $\mathcal{K}_{11} = -8.4$~$\mu$eV, $\mathcal{K}_{22} = -1.11$~$\mu$eV, and $\mathcal{K}_{12} =  -3.2$~$\mu$eV, representing 5-8\% of the corresponding bilinear strengths. Remarkably, the Co1-Co1 bonds exhibit exceptionally large DMI with $|\mathcal{D}| = 0.099$~meV yielding $|\mathcal{D}/\mathcal{J}| \simeq 0.84$, atypical for insulating chiral oxides, while Co2-Co2 bonds show negligible DMI ($|\mathcal{D}/\mathcal{J}| \simeq 0.03$) and Co1-Co2 bonds display intermediate values ($|\mathcal{D}/\mathcal{J}| \simeq 0.102$). As schematically shown in Fig.~\ref{Fig1}B, the calculated SIA constants are site-dependent with $\mathcal{A}_1 = -2.1$~meV (easy-axis along [111] for Co1 at 12$d$ sites) and $\mathcal{A}_2 = 3.1$~meV ($xy$-easy-plane for Co2 at 8$c$ sites), representing the dominant energy scale, approximately 10-15 times larger than exchange interactions. In the absence of strong anisotropy in a single sublattice system without interaction, the canonical DMI helimagnet relation $\lambda_\text{h} = 4\pi\mathcal{J}a/\mathcal{D}$, predicts $\lambda_\text{h} \simeq 5.3$~nm for Co1-Co1, coincidentally close to the observed range of 5.7-9.6~nm across the temperature-dependent phases. The same relation predicts $\lambda_\text{Co1-Co2} \simeq 45$~nm and $\lambda_\text{Co2-Co2} \simeq 109$~nm, both incompatible with experimental observations. The calculated energy hierarchy $\mathcal{A} \gg \mathcal{J} > \mathcal{D} \gg \mathcal{K}$, combined with competing antiferromagnetic exchanges on the two-sublattice spinel-like structure, identifies \ce{Co5TeO8} as a frustrated magnetic system in which the SIA exceeds the exchange and DMI. In this regime, frustration and DMI drive the underlying finite-$\bm{Q}$ instability and chirality, while the dominant site-dependent SIA is proposed to lift the directional degeneracy and to favor a propagation vector selected from the otherwise nearly degenerate manifold.
	
	This counterintuitive strong SIA can be traced to the electronic structure of high-spin Co$^{2+}$ ($d^7$, $S = 3/2$) in distinct local environments. For the octahedral Co1 sites (12$d$, $t_{2g}^5e_g^2$), the incompletely filled $t_{2g}$ manifold is expected to retain unquenched orbital angular momentum, in contrast to either fully filled or half-filled configurations. The Co1 octahedron exhibits a local $\langle\text{111}\rangle$-oriented distortion where the six Co-O coordinations comprise two short (2.01~\AA), two intermediate (2.14~\AA), and two long (2.17~\AA) bond lengths, with the two short bonds subtending an expanded angle of 97.3$^\circ$ and the two long bonds subtending a contracted angle of 86.3$^\circ$, consistent with compression along the pseudo-three-fold axis. This distortion splits the parent $^4T_{1g}$ term into $^4A_{2g}$ and $^4E_g$ components in $D_{3d}$ symmetry. Mixed with spin-orbit coupling, this split manifold produces zero-field splitting of the $S = 3/2$ states with easy-axis character along the compression direction, yielding in our calculations $\mathcal{A}_1 = -2.1$~meV~\cite{gransbury2019single,tripathi2021structure}. In contrast, the tetrahedrally-coordinated Co2 (8$c$ sites, $e^4_gt^3_{2g}$) presents a distinct scenario. The $^4A_2$ ground term of tetrahedral Co$^{2+}$ is an orbital singlet with no first-order orbital angular momentum. Nevertheless, Co2 occupies a trigonally compressed tetrahedron ($C_{3v}$ symmetry) with one short Co-O bond (1.91~\AA) along the three-fold axis and three longer bonds (2.02~\AA), yielding O-Co-O angles of 106.6$^\circ$ and 112.2$^\circ$. This distortion modifies both the energies and symmetries of the excited-state manifold (primarily $^4T_2$) that couples to the ground state via spin-orbit interaction. Our \textit{ab initio} calculations indicate that this second-order mechanism produces easy-plane anisotropy ($\mathcal{A}_2 = 3.1$~meV), whose magnitude appears to be enhanced by covalent mixing with heavy tellurium ligands. This overlap is expected to increase the effective spin-orbit coupling at Co2 sites and amplifies these second-order contributions despite the absence of first-order orbital effects. The insulating state with persistent correlations in \ce{Co5TeO8} is likely essential for preserving this strong SIA. In itinerant systems, electron delocalization into broad conduction bands quenches orbital angular momentum and averages over local distortions, substantially reducing magnetic anisotropy. This suppression is evidenced by the fourfold reduction in near-metallic \ce{CoV2O4} compared to the wider-gap isostructural Mott insulator \ce{MnV2O4}, where electron localization preserves the orbital physics~\cite{lee2017magnetic,kismarahardja2011co,katsufuji2020nonlinear}. The combination of Te-induced distortions at both Co sites, unquenched orbital moments at Co1, heavy-ligand spin-orbit enhancement at Co2, and Mott localization preserving the local electronic structure is proposed to produce the dominant site-dependent SIA.
	
	
	\begin{center}
		\textbf{DISCUSSION}
	\end{center}
	
	\begin{table*}[htb!]
		\begin{center}
			\caption{\textbf{List of non-centrosymmetric magnets hosting incommensurate spin textures.}  Here, DMI and RKKY refer to Dzyaloshinskii-Moriya interaction and Ruderman–Kittel–Kasuya–Yosida interaction, respectively. ``DMI-based'' refers to where DM interaction competes with leading order exchange to stabilize the incommensurate ground state. Superscripted ``c'' stands for the electrically conducting members from the list of compounds. Note that some systems, such as \ce{EuNiGe3} may possess both Bloch and Néel-type twisting, aptly designated by ``mixed'' helicity.}
			\label{tab:TopoList}
			\centering
			\begin{tabular}{|c|c|c|c|c|c|}
				\Xhline{2\arrayrulewidth}
				\Xhline{2\arrayrulewidth}
				
				\multirow{2}{6em}{\centering Material} &
				\multirow{2}{4em}{\centering Space group} & 
				\multirow{2}{4em}{\centering Pitch length} & 
				\multirow{2}{6em}{\centering Helicity} & \multirow{2}{4em}{\centering Stabilization mechanism} & \multirow{2}{2em}{\centering Ref.} \\
				
				&&&&& \\
				\Xhline{2\arrayrulewidth}
				\Xhline{2\arrayrulewidth}
				
				\ce{EuNiGe3}$^\text{c}$ & $I4mm$ (107) & 1.7~nm & Mixed & RKKY + DM interaction & \cite{singh2023transition} \\ \hline
				
				\ce{EuPtSi}$^\text{c}$ &$ P2_{1}3$ (198) & 1.8~nm & Bloch & Geometric frustration + DM & \cite{kaneko2019unique} \\
				& & & & + RKKY + four-spin interactions & \\ \hline
				
				\ce{MnGe}$^\text{c}$ & $P2_{1}3$ (198) & 2.8-5.5~nm & Bloch & Exchange frustration & \cite{kanazawa2012possible} \\ \hline
				
				\ce{YbNi3Al9}$^\text{c}$ & $R32$ (155) & 3.4~nm & Bloch & RKKY+DMI & \cite{matsumura2017chiral} \\ \hline
				
				\ce{Ba3NbFe3Si2O14} & $P321$ (150) & 3.7~nm & Bloch & Exchange frustration & \cite{marty2008ba3nbfe3si2o14} \\ \hline
				
				\Xhline{2\arrayrulewidth} 
				
				\textbf{\ce{Co5TeO8}} & $\bm{P4_{3}32}$/$\bm{P4_{1}32}$ & \textbf{5.7-10~nm} & \textbf{Bloch} & \textbf{SIA + DMI} & \textbf{This} \\ 
				& \textbf{(212/213)} & & & \textbf{+ frustration} & \textbf{work} \\
				\hline \Xhline{2\arrayrulewidth}
				
				\ce{CeAlGe}$^\text{c}$ & $I4_1md$ (109) & 6~nm & Néel & RKKY + DMI & \cite{puphal2020topological} \\ \hline
				
				\ce{Y3Co8Sn4}$^\text{c}$ & $P6_3mc$ (186) & 8-16~nm & In-plane vortex & four-spin interaction & \cite{takagi2018multiple} \\ \hline
				
				\ce{Pr5Ru3Al2}$^\text{c}$ & $I2_13$ (199) & 15~nm & Bloch & DMI-related & \cite{makino2016incommensurate} \\ \hline
				
				\ce{Ni2InSbO6} & $R3$ (146) & 15.7~nm & Bloch & DMI-based & \cite{araki2020metamagnetic} \\ \hline
				
				\ce{MnSi}$^\text{c}$ & $P2_{1}3$ (198) & 17~nm & Bloch &	DMI-based & \cite{muhlbauer2009skyrmion}\\ \hline
				
				\ce{Ba2CuGe2O7} & $P\bar{4}2_1m$ (113) & 22~nm & Bloch &	DMI-based & \cite{muhlbauer2012phase} \\ \hline
				
				\ce{GaV4(S/Se)8} & $ R3m $ (160) & 22~nm & Néel &	DMI-based & \cite{kezsmarki2015neel} \\ \hline
				
				\ce{CrTa3S6}$^\text{c}$ & $P6_322$ (182) & 22.5~nm & Bloch & DMI-based & \cite{kousaka2016long} \\ \hline
				
				Fe$ _{1-x} $Co$ _{x} $Si$^\text{c}$ & $P2_{1}3$ (198) & 30-200~nm & Bloch &	DMI-based & \cite{grigoriev2009crystal} \\ \hline

				\ce{Cr_{1/3}NbS2}$^\text{c}$ & $P6_322$ (182) & 48~nm & Bloch &	DMI-based & \cite{togawa2012chiral} \\ \hline
				
				\ce{Cu2OSeO3} & $P2_{1}3$ (198) & 62~nm & Bloch &	DMI-based & \cite{adams2012long} \\ \hline
				
				\ce{BiFeO3} & $R3c$ (161) & 62~nm & Néel & DMI-based & \cite{sosnowska1982spiral} \\ \hline
				
				\ce{FeGe}$^\text{c}$ & $P2_{1}3$ (198) & 70~nm & Bloch &	DMI-based & \cite{yu2011near} \\ \hline
				
				Co$ _{x} $Zn$ _{y}$Mn$ _{z}$$^\text{c}$ & $P4_{3}32$/$P4_{1}32$ (212/213) & 70-130~nm & Bloch & DMI-based & \cite{tokunaga2015new} \\ \hline
				
				FeCo$ _{0.5} $Rh$ _{0.5} $Mo$ _{3} $N & $P4_{1}32$ (213) & 110~nm & Bloch & DMI-based & \cite{li2016emergence} \\ \hline
				
				\ce{NdFe3(BO3)4} & $R32$ (155) & 114~nm & Bloch & DMI-based & \cite{janoschek2010single} \\ \hline
				
				\ce{VOSe2O5} & $ P4cc $ (103) & 125-180~nm & Néel &	DMI-based & \cite{kurumaji2017neel} \\ \hline
				
				Mn$ _{1.4} $Pt$ _{0.9} $Pd$ _{0.1} $Sn$^\text{c}$ & $ I\bar{4}2m $ (121) & $>$~130~nm & Mixed & Dipolar + anisotropic & \cite{nayak2017magnetic} \\
				& & & & DM interaction & \\ \hline 
				
				Fe$_{1.9}$Ni$_{0.9}$Pd$_{0.2}$P$^\text{c}$ & $ I\bar{4}2m $ (121) & 280~nm & Mixed & Dipolar + anisotropic & \cite{karube2021room} \\
				& & & & DM interaction & \\
				
				\hline\Xhline{2\arrayrulewidth} \Xhline{2\arrayrulewidth}
			\end{tabular}
		\end{center}
	\end{table*}

	The microscopic origin of sub-10~nm helimagnetism in \ce{Co5TeO8} presents a compelling materials physics puzzle: in absence of strong anisotropy, canonical DMI helimagnets achieve $\lambda_\text{h} \sim$ 18-130~nm, yet \ce{Co5TeO8} realizes $\lambda_\text{h} \sim$ 5.7–9.6~nm despite only moderately enhanced $|\mathcal{D}/\mathcal{J}|$ compared to archetypal chiral magnets [e.g., MnSi $\sim$ 0.12~\cite{shanavas2016electronic}; \ce{Cu2OSeO3} $\sim$ 0.04–0.58~\cite{janson2014quantum}]. Our \textit{ab initio} calculations point to a different mechanism in which site-dependent SIA, an order of magnitude larger than all exchange interactions, provides the leading energy scale. Three competing antiferromagnetic exchanges on the two-sublattice structure create geometric frustration, producing a nearly flat $E(\bm{Q})$ landscape over 0.8-1.3~nm$^{-1}$. Within this degenerate manifold, dominant SIA is proposed to select the helical wavevector through $\bm{Q}$-dependent orientational averaging, where shorter wavelengths reduce anisotropy penalties by allowing larger inter-site spin canting that better accommodates the conflicting easy-axis (Co1) and easy-plane (Co2) local environments. While the canonical $\lambda_\text{h}$ = $4\pi\mathcal{J}/\mathcal{D}$ relationship using Co1–Co1 parameters coincidentally predicts $\lambda_\text{h} \sim$ 5.2~nm, this numerical agreement may mask distinct physics: large DMI creates the helical instability and establishes chirality, but we argue that SIA provides the leading contribution to  wavelength selection within the frustrated landscape. This SIA-selection mechanism is supported by $T$- and $H$-dependent neutron scattering data, as detailed below. Additionally, weak ferromagnetic-type biquadratic interactions (5–8\% of bilinear strength) lift residual degeneracies within the frustrated $\bm{Q}$-space manifold and provide energy barriers between competing helical variants, enabling sharp phase transitions and hysteretic behavior. Although individual exchange bonds are modest, the experimentally observed $T_\text{HM}$ is driven by collective stabilization from the full frustrated lattice and strong anisotropy, as captured by mean-field analysis (Supplementary Materials Section~{\color{Red}S5}). This mechanism is intrinsically tied to the predicted insulating ground state with strong electronic correlations, which favors the localized orbital physics underlying the enhanced SIA while maintaining sufficient correlation strength to generate competing exchange pathways. In summary, \ce{Co5TeO8} thus appears to realize a DMI-initiated, SIA-enhanced, frustration-enabled compact helimagnet, and may serve as a prototype for a class of anisotropy-dominated helimagnets in strongly correlated insulators.
	
	Another indication of SIA dominance emerges from the field-dependence of the helical wavevector. \ce{Co5TeO8} exhibits field-driven compression ($d\bm{Q}/dH >$ 0 as shown in Fig.~\ref{Fig5}F), not what is expected from canonical DMI helimagnets for which the modulation period is typically field independent to first approximation. In \ce{Co5TeO8}, the field reorients the spin configuration, and on the frustrated two-sublattice landscape this favors shorter helical pitches that better accommodate the conflicting easy-axis (Co1) and easy-plane (Co2) environments. This distinctive field response is difficult to reconcile with a purely DMI-controlled wavelength selection and is most naturally explained if SIA provides the leading contribution. The synergy between frustration and SIA manifests dramatically in high-field magnetization, which fails to saturate even at 60 T ($E_\text{Zeeman} \sim 10.4$~meV), as shown in Fig.~\ref{Fig2}E. This non-saturation arises from geometric frustration creating competing exchange fields that reduce the effective molecular field, while calculated SIA barriers ($\sim$24~T for Co1, $\sim$36~T for Co2) must be overcome to fully polarize spins. The contrasting anisotropy character of the two sublattices, combined with competing interactions, underlies the rich eight-phase magnetic diagram, where phase boundaries reflect competition between helical wavevector selection, rotation plane orientation, and Co1–Co2 sublattice coupling. Further experiments on \ce{Co5TeO8} single crystals combined with theoretical modeling are required to validate these points.
	
	The SIA-dominated mechanism is further supported by two distinct temperature-induced features. First, at $T_\text{HM}$, long-range helical order emerges with a temperature-dependent wavevector $|\bm{Q}(T)|$ that evolves as thermal fluctuations partially quench the effective anisotropy. Second, upon cooling through a first-order lock-in transition at $T_\text{lock}$ (= $T_\text{hys}^{(2)} \sim 25$~K) exhibiting substantial thermal hysteresis (Fig.~\ref{Fig2}A), the helical period becomes temperature-independent and rigidly fixed at $\lambda_\text{h}$ = 5.7~nm. The energy scale $k_\text{B}T_\text{lock} \sim 2.15$~meV is comparable to the calculated Co1 single-ion barrier $\mathcal{A}_1$ (also naturally lower than $\mathcal{A}_2$), suggesting that reduction of thermal energy below this scale may trigger the transition, and pins the system in the SIA-optimized configuration. This two-transition phenomenology bears striking similarity to the archetypal Ising chain material \ce{CoNb2O6}, which also exhibits an incommensurate–commensurate lock-in~\cite{heid1995magnetic}, but with a fundamental mechanistic difference. In \ce{CoNb2O6}, ultra-strong crystal-field splitting ($\sim$30~meV) quenches the spin to $S_\text{eff}$ = 1/2 and rigidly locks spin directions along a single axis, permitting only amplitude modulation of collinear ferromagnetic chains in the intermediate phase; lock-in is then driven by weak inter-chain coupling, occurring at an energy scale much smaller than the crystal-field splitting. In contrast, \ce{Co5TeO8}'s moderate SIA preserves $S = 3/2$ and permits continuous spin rotation, resulting in helimagnetic spiral, with lock-in occurring at $T_\text{lock} \sim \mathcal{A}_1/k_\text{B}$ via anisotropy saturation rather than interchain coupling. The hysteretic first-order character in \ce{Co5TeO8} points to an energy landscape with multiple metastable $\bm{Q}$-states separated by nucleation barriers, consistent with the frustrated $E(\bm{Q})$ manifold. Above $T_\text{lock}$, the system occupies a thermally accessible ``soft'' $\bm{Q}$-state with broader distribution, while below $T_\text{lock}$, it nucleates into a ``locked'' $\bm{Q}$-state with narrow, rigid distribution. This comparison supports the picture of SIA-dominated helical behavior with sufficient DMI and frustration in \ce{Co5TeO8}, fundamentally distinct from DMI-controlled systems with monotonic changes in helical period~\cite{baral2023direct}.
	
	The calculated anisotropy in \ce{Co5TeO8} places it in an intermediate regime within the broader landscape of chiral magnetism. At one extreme, archetypal cubic chiral magnets with weak magnetocrystalline anisotropy preserve near-degeneracy among multi-$\bm{Q}$ propagation vectors, enabling continuous $\bm{Q}(T)$ evolution and field-stabilized skyrmion lattices; metallic MnSi, Fe$_{1-x}$Co$_x$Si, and the insulating \ce{Cu2OSeO3} exemplify this regime. At the opposite extreme, \ce{CoNb2O6}'s ultra-strong anisotropy restricts magnetic modulation to amplitude fluctuations of collinear order, suppressing all non-collinear textures and enforcing rigid uniaxial Ising behavior. \ce{Co5TeO8} lies between these limits where site-dependent SIA dominates exchange by at least an order of magnitude, yet remains small enough in absolute terms to preserve unquenched $S = 3/2$ and permit continuous spin rotation. Unlike weak-anisotropy systems where higher-harmonic coupling enables multi-$\bm{Q}$ states~\cite{leonov2015multiply}, the dominant SIA in \ce{Co5TeO8} strongly penalizes spin textures that deviate from locally optimized easy axes defined by the chiral crystal structure. A multi-$\bm{Q}$ superposition, which inherently requires spins to point in nearly isotropic directions to satisfy multiple wavevectors, incurs significantly higher anisotropy cost compared to a single-$\bm{Q}$ helix that can coherently adapt its phase to the local anisotropy axes~\cite{hayami2021noncoplanar}. Consequently, the system is driven toward a single-$\bm{Q}$ harmonic state, effectively ``locking'' the magnetic texture to the structural chirality and suppressing complex multi-$\bm{Q}$ topologies. The lock-in transition and field-driven helix compression described above are absent in weaker-anisotropy systems, consistent with SIA playing the leading role. The $\beta$-Mn-type \ce{Co8Zn8Mn4}, with comparable DMI but weaker anisotropy, provides direct comparison, as it exhibits a longer $\lambda_\text{h}$ (Table~\ref{tab:TopoList}) while retaining skyrmion phases, supporting the role of SIA in suppressing multi-$\bm{Q}$ skyrmion phases~\cite{tokunaga2015new}. These findings suggest a transferable design principle: systematically weakening SIA through selective dilution or enhancing DMI via heavier 4$d$/5$d$ element substitution at Co$^{2+}$ sites could generate skyrmion derivatives in isostructural compounds. More broadly, \ce{Co5TeO8} exemplifies how strong single-ion terms in frustrated geometries expand the accessible magnetic texture space beyond DMI paradigms, motivating exploration of anisotropy-engineered noncollinear states in other frustrated pyrochlore and spinel oxides with heavy transition metals. We note, however, that the present polycrystalline measurements do not permit a direct experimental determination of the full anisotropy tensor; accordingly, the microscopic interaction values should be viewed as semi-quantitative estimates whose main significance lies in the inferred energy-scale hierarchy.
	
	Our work establishes \ce{Co5TeO8} as a magnetic insulator hosting a short-period helical ground state, with calculations indicating the presence of strong electronic correlations. The stabilization mechanism is proposed to arise from a distinctive competition between site-dependent SIA, DMI and frustration, unlike that seen in archetypal cubic helimagnets. Our results suggest a design principle that links the local coordination geometry to the nanoscale spin texture. The interplay between SIA and DMI proposed here broadens the search for short-period helimagnets beyond compounds stabilized by DMI alone, extending it to material classes with weak to moderate anisotropy. Mixed-coordination structures, such as Co-based chalcogenide spinels and $\beta$-Mn-type compounds, are therefore natural candidates for systematic exploration of this mechanism. The short-period helical texture itself may find practical use in three areas. First, in high-frequency nano-oscillators, where the insulating ground state avoids eddy-current losses and the antiferromagnetic background cancels the Magnus force, both favoring dynamical stability~\cite{gupta2025ultrahigh}. Second, in reservoir computing, where the two crystallographically distinct Co sites could provide additional nonlinear degrees of freedom compared with single-sublattice magnetic domains~\cite{lee2024task}. Third, in magneto-electric switching, where the insulating ground state allows electric-field rather than current-driven control---an advantage for device miniaturization, contingent on sufficiently large magneto-electric tensor elements in \ce{Co5TeO8}~\cite{mostovoy2024multiferroics}. Quantitative measurement of the Mott gap and the magneto-electric coefficients will therefore be needed to assess the practical viability of voltage-controlled manipulation of the helices. Further experiments to resolve the magnon band structure of these compact helices~\cite{che2025short}, to probe nonreciprocal thermal transport arising from broken inversion symmetry~\cite{ding2025intrinsic}, and to measure electric-field-driven phase transitions~\cite{stoliar2013universal} would clarify the functional potential of \ce{Co5TeO8} and related materials for low-dissipation spintronics. More broadly, this work suggests that competing anisotropies between inequivalent magnetic sites, usually regarded as detrimental to magnetic order, can instead serve as a design principle for engineering compact spin textures in correlated insulators.

	
	
	\begin{center}
		\textbf{METHODS}
	\end{center}
	
	\noindent{\bf Density functional theory calculations}
	
	\noindent Density functional theory calculations were performed using the Vienna {\it Ab-initio} Simulation Package (VASP)\cite{kresse1996vasp}. Electron-core interactions are described using the projector-augmented wave method\cite{bloechl1994paw}, while the Kohn-Sham wavefunctions for the valence electrons were expanded in a plane wave basis with a cut-off on the kinetic energy of 400~eV. LDA+$U$ calculations were performed within the local density approximation framework with a Coulomb on-site repulsion $U= 3$~eV acting on the $d$ shell of Co atoms following the rotationally invariant scheme proposed by Dudarev \cite{dubarev1998electron}. The calculations included spin-orbit coupling and non-collinear magnetism to capture essential magnetism of \ce{Co5TeO8}. The cubic cell ($a = 8.535$~\AA) had 56 atoms, and the Brillouin zone sampling used 100 $k$-points with the Methfessel-Paxton smearing (width 0.01~eV). The intrinsic $C_3$ symmetry of the system was utilized, and static single-point energy calculations were performed without ionic relaxation. It should be noted that residual self-interaction errors inherent to LDA often result in slight underestimation of band gaps in systems with localized 3$d$ electronic states. Thus, we may expect a slightly larger band gap in reality.
	
	\vspace{5mm}
	
	{\noindent\bf Microcrystalline sample preparation}
	
	\noindent The pure phase microcrystalline \ce{Co5TeO8} sample was synthesized using conventional solid-state reactions. Stoichiometric amounts of \ce{CoCO3} (Aldrich, 99.999\%) and \ce{TeO2} (Aldrich, 99.999\%) were thoroughly mixed using a mortar. The resultant mixture was transferred to a platinum crucible and placed inside a high-temperature muffle furnace. The temperature of the furnace was slowly raised to 1000 $^\circ$C at a rate of 100 $^{\circ}$C per hour. The furnace was then dwelled for an hour at that temperature. Followed by increasing the temperature to 1100 $^\circ$C at slower rate of 50 $^\circ$C per hour. The powder was then sintered at this temperature for 72 hours, followed by quenching to room temperature. The quenching process was necessary to avoid co-crystallization of \ce{Co3O4} spinel. Results presented throughout this article were obtained on the same batch of microcrystalline sample.
	
	\vspace{5mm}
	
	{\noindent\bf Single crystal X-ray diffraction}
	
	\noindent Microcrytalline \ce{Co5TeO8} grains with suitable size were mounted on to the goniometer head with cryo-loop obtained from MiTeGen, followed by installing on a Rigaku Synergy-R single crystal diffractometer, equipped with Cu/Mo Rotating-anode X-ray source and an Atlas CCD Detector. The temperature of the setup was controlled by an Oxford Cryostream 800 Plus system.
	
	\vspace{5mm}
	
	\noindent{\bf Electron backscatter diffraction (EBSD) and energy-dispersive X-ray spectroscopy (EDX) measurements}
	
	\noindent We performed two EBSD and EDX experiments. For the first experiment, \ce{Co5TeO8} particles were distributed over a silver-pasted stub, and EBSD spectra were collected using a field emission gun (FEG) scanning electron microscope (SEM) Zeiss Gemini 450 equipped with Oxford Instruments UltimMax EDX detector and CMOS EBSD Symmetry camera. Data acquisition was performed with the AZtec system, and subsequent processing utilized the AZtecCrystal software suite. In the second experiment, \ce{Co5TeO8} particles were embedded in resin and cured at 180 $^\circ$C for 5 minutes. Prior to EBSD and EDX analysis on the same specimen, a thin carbon layer was deposited in order to reduce the electron irradiation-induced charging effects.
	
	\vspace{5mm}
	
	{\noindent\bf Magnetometry, heat capacity, and capacitance measurements}
	
	\noindent DC magnetization as well as AC susceptibility measurements were performed using the VSM and ACMS-II option of a commercial Quantum Design (QD) 14~T Physical Property Measurement System (PPMS), as well as a Magnetic Property Measurement System (MPMS), MPMS3 from QD. The polycrystalline sample was enclosed in a standard polypropylene holder provided by QD. High-field magnetization experiments were performed in pulsed magnetic fields up to $\mu_0$H = 60~T at the Dresden High Magnetic Field Laboratory, which were normalized to laboratory measurements up to $\mu_0$H = 14 T.
	
	Specific heat measurements were performed on a pelletized sample. In order to produce a pellet with higher structural stability, microcrystalline sample was first crushed in a mortar for about 5~minutes. The diameter and height of the disc-shaped pellet were made to be 3~mm and 0.5~mm, respectively. Measurements were performed using the Heat Capacity option of the 14~T PPMS, in the typical relaxation method. The pellet was mounted on the 3 $\times$ 3~mm$^2$ sapphire plate provided by QD. Apiezon-N grease was used to ensure good thermal contact between the sample and the platform. Later, the long-pulse technique was also used to measure the sample heat capacity.
	
	Capacitance measurements were performed on a pressed \ce{Co5TeO8} pellet of size 5 $\times$ 5 $\times$ 0.5 mm, using an AH2700A ultraprecision capacitance bridge in zero dc field, and with a 1~kHz excitation and a voltage of 15 V. The sample mount was installed onto 9~T PPMS. Capacitance measurements were performed as a function of magnetic field at fixed temperatures in a sweep mode.
	
	\vspace{5mm}
	
	\noindent{\bf Electron energy loss spectroscopy (EELS) experiment}
	
	\noindent The experimental determination of the cation oxidation state in \ce{Co5TeO8} was carried out on a thin lamella prepared from a microcrystalline specimen using focused ion beam (FIB). We used the double aberration corrected, monochromated Titan Themis at the Interdisciplinary Centre for Electron Microscopy (CIME) of \'Ecole Polytechnique F\'ed\'erale de Lausanne (EPFL). We acquired spectrum images (see Supplementary Materials section) in the dual EELS mode, which allowed for calibrating the absolute edge energy position and determining chemical shifts in the edge energies associated with changes in valence. We chose to acquire spectra at dispersions of 0.1 eV/channel and dewell time of 50~ms in an which provided high energy resolution and produced well-defined \textit{L}-edges above background in a single pixel-spectrum but limited the electron beams interactions and radiation damage.
	
	\vspace{5mm}
	
	{\noindent\bf Neutron diffraction experiments}
	
	\noindent Wide-angle neutron powder diffraction measurements were carried out on a polycrystalline \ce{Co5TeO8} sample using the high-resolution powder diffractometer (HRPT) of the SINQ instrumental suite at Paul Scherrer Institute (PSI). Data were collected in both medium-resolution and high-resolution modes, employing neutrons with a fixed wavelength of 2.45~\AA. Each measurement was preceded by a 10-minute thermal stabilization period to ensure equilibrium, followed by a 2-hour acquisition of high-statistics data.
	
	\vspace{5mm}
	
	{\noindent\bf Small angle neutron scattering (SANS) experiments}
	
	\noindent SANS experiments, without polarization-analysis, were carried out using the SANS-I instrument of Swiss Spallation Neutron Source (SINQ) located at the Paul Scherrer Institute (PSI). About 200~mg microcrystalline sample was mixed thoroughly with a 1:1 mixture of hydrogen-free FC-77 and FC-75 Fluorinert liquid and enclosed in a vanadium can. The Fluorinert freezes upon cooling to cryogenic temperatures, providing a quasi-strain-free matrix that maintains the random zero-field orientations of the sample grains in finite magnetic field. The can was then installed inside a 6.8 T horizontal field magnet. For all of our experiments, neutrons with wavelength 5 \AA~with a FWHM spread $\Delta\lambda/\lambda$ = 10\% were used. The incoming neutron beam was collimated by a distance of 8~m before the sample, whereas the scattered neutrons were collected by a two-dimensional $^3$He multidetector placed 3.5~m behind the sample.
	
	Uniaxial polarization analysis was performed using the D33 small-angle neutron scattering instrument at the Institut Laue-Langevin (ILL), Grenoble. The polycrystalline \ce{Co5TeO8} sample was embedded in hydrogen-free glue and encapsulated within a vanadium can. Measurements utilized longitudinal polarization geometry with the incident neutron beam collinear to the applied magnetic field. A weak longitudinal magnetic field of 50~mT was applied to the sample, while a 10~mT guide field maintained the neutron spin quantization axis parallel to the incident beam. Scattered neutrons were analyzed using a spin-polarized $^3$He cell with maximum polarization of 98.6\%. Four neutron spin states ($\downarrow\downarrow$, $\uparrow\downarrow$, $\uparrow\uparrow$, and $\downarrow\uparrow$) were systematically measured as a function of temperature during cooling. Neutron beam depolarization was monitored at two-hour intervals in the paramagnetic regime ($T = 65$~K). Data analysis was conducted using GRAS$_{\text{ANS}}$P software~\cite{dewhurst2023graphical}.
	
	\vspace{5mm}
	
	\noindent{\bf Cluster-based quantum chemistry calculations} 
	
	\noindent We performed many-body wavefunction calculations on electrostatically embedded finite-size cluster models extracted from the experimental crystal structure. The crystallographic unit cell contains two distinct Co$^{2+}$ sites: Co1 atoms in distorted octahedral coordination and Co2 atoms in tetrahedral coordination with oxygen ligands, as shown in Fig.~\ref{Fig1}A. Our computational models feature a central cluster containing one or two coordination polyhedra, with the active Co$^{2+}$ centers treated using multireference wavefunctions. Adjacent Co polyhedra representing nearest-neighbor environments were included at the Hartree-Fock level to account for local charge distribution effects. Long-range electrostatic interactions were incorporated through an array of point charges fitted to reproduce the crystalline Madelung potential~\cite{klintenberg2000accurate}. Electron correlation effects in the central clusters were described using complete-active-space self-consistent-field (CASSCF) and subsequent multi-reference configuration interaction (MRCI) calculations. The MRCI treatment incorporated single and double excitations from the Co$^{2+}$ 3$d$ valence orbitals and the bridging oxygen 2$p$ orbitals. Co$^{2+}$ oxidation states were confirmed through electron energy loss spectroscopy measurements (see Supplementary Materials Section~{\color{red}2}), providing the electronic configuration basis for all quantum chemical calculations. The multiplet structure obtained then entered into spin-orbit calculations to obtain spin-orbit coupled eigenstates and wavefunctions~\cite{helgaker2014molecular}.
	
	The generalized spin Hamiltonian considered to determine the local magnetic interactions in \ce{Co5TeO8} can be expressed as
	
	\begin{equation}
		\mathcal{H} = \mathcal{H}^\text{I} + \mathcal{H}^{\text{II}},
		\label{eqn:TotalHamil}
	\end{equation}
	
	\noindent where $ \mathcal{H}^{\text{I}} $ denotes the coupling between the $i^{\mathrm{th}}$ and $j^{\mathrm{th}}$ nearest-neighbor sites bearing spin ${\mathcal{S}}_i$ and ${\mathcal{S}}_j$, respectively, while $ \mathcal{H}^\text{II} $ contains the contribution of the interactions within sites of spin ${\mathcal{S}}_i$. The inter-site contribution, $\mathcal{H}^{\text{I}}$ reads as

	\begin{equation}
		\mathcal{H}^{\text{I}} =  \mathcal{H} =  \sum_{\langle i,j\rangle = 1,2}\mathcal{J}_{ij} (\mathcal{S}_i\cdot \mathcal{S}_j) + \mathcal{K}_{ij}(\mathcal{S}_i \cdot {\mathcal{S}_j})^2 + \mathcal{D}_{ij} \cdot \left({\mathcal{S}_i}\times {\mathcal{S}_j}\right)
		\label{eqn:HamilExchange}
	\end{equation}

	\noindent with $\mathcal{J}_{ij}$ and $\mathcal{K}_{ij}$ being the bilinear and biquadratic isotropic exchange couplings, respectively, and $\mathcal{D}_{ij}$ is the antisymmetric DMI vector~\cite{dzyaloshinsky1958thermodynamic,moriya1960anisotropic}. The intra-site contribution $\mathcal{H}^\text{II}$ takes the following form:
	
	\begin{equation}
		\mathcal{H}^\text{II}  = \sum_{i}\mathcal{A}_i(\mathcal{S}_{i}\cdot \hat{e}_i)^2.
		\label{eqn:HamilSIA}
	\end{equation}
	
	\noindent Here, $\mathcal{A}_i$ is the single-ion anisotropy. Further details about our quantum chemistry calculations can be found in the Supplementary Materials Section~{\color{red}5}.

	
	\vspace{10pt}
	
	\begin{center}
		\textbf{REFERENCES}
	\end{center}
	
	
	\bibliography{bib}
	\bibliographystyle{sciencemag}
	
	
	\vspace{10pt}
	
	\noindent\textbf{Acknowledgments}: P.R.B. and J.S.W. thank C.~Pfleiderer, S.~Seki, S.~Hayami, and O. Utesov for fruitful discussions. We thank C. Cayron, B. Klemke, M. Bonnaud and O. Zaharko for their help with EBSD, polarization, D33-SANS, and pulsed magnetic field measurements, respectively.
	
	\vspace{5pt}
	
	\noindent\textit{Funding:}
	P.R.B. acknowledges Swiss National Science Foundation (SNSF) Postdoc.Mobility grant P500PT\_217697 and Return CH Postdoc.Mobility grant P5R5-2\_239356 for financial support. Furthermore, the following SNSF projects are acknowledged: 200021\_188707 (P.R.B., V.U., \& J.S.W.), 200020\_182536 (P.R.B.), and Sinergia Network ``NanoSkyrmionics'' grant no. CRSII5\_171003 (P.R.B., R.Y., V.U., H.M.R., O.V.Y., A.M., \& J.S.W.). We acknowledge beamtime allocations from PSI (20200295, 20200297, 20221263, 20222723, 20230476), ILL (5-41-1226) and HLD-HZDR (member of the European Magnetic Field Laboratory (EMFL)) facilities. This work is based partly on experiments performed at the Swiss spallation neutron source SINQ, Paul Scherrer Institute, Villigen, Switzerland. Computations were performed at the Scientific IT and Application Support Center (SCITAS) of EPFL.
	
	\vspace{5pt}
	
	\noindent\textbf{Author contributions}:
	P.R.B. and A.M. grew the micro-crystalline sample. Single crystal diffraction experiments were performed and analyzed by W.H.B with extensive guidance from A.M. Magnetometry and specific heat measurements were performed by P.R.B and I.\v{Z} , whereas M.B. performed the capacitance measurements. EELS measurements were performed by P.R.B. and T.L. Wide-angle neutron diffraction was performed by P.R.B., V.U., V.P., H.M.R., \& J.S.W. P.R.B., V.U., H.M.R \& J.S.W. performed the non-polarized SANS while, polarized SANS experiment was performed by P.R.B, V.U., R.C., and N.J.S. P.R.B, and Y.S. performed the high-field magnetization experiments. {\it Ab initio} many-body wavefunction calculations as well as density functional theory calculations were performed by R.Y., with support from O.V.Y. P.R.B. \& J.S.W wrote the manuscript, with input from R.Y. All authors read and commented on the manuscript. P.R.B., O.V.Y., A.M. and J.S.W. jointly conceived the project.
	
	\vspace{5pt}
	
	\noindent\textbf{Competing interests}:
	There are no competing interests to declare.
	
	\vspace{5pt}
	
	\noindent\textbf{Data and materials availability}:
	Small angle neutron scattering data obtained at D33 can be found at DOI: {\color{blue}link}. All other data can be found in the Zenodo online repository with unique identifier: {\color{blue}link}.

\end{document}